\documentclass[12pt]{JHEP}
\input colordvi

\title{{\textRed Solar neutrinos: global analysis and implications for
SNO}\textBlack} \author{John N. Bahcall\footnote{E-mail address:
jnb@ias.edu}\\ School of Natural Sciences, Institute for Advanced
Study, Princeton, NJ 08540} \author{Plamen I. Krastev\thanks{E-mail
address: krastev@ias.edu}\\ School of Natural Sciences, Institute for
Advanced Study, Princeton, NJ 08540,\\ Dept. of Physics and Astronomy,
University of Delaware, Newark, DE 19716} \author{Alexei
Yu. Smirnov\thanks{E-mail address: smirnov@ictp.trieste.it}\\
International Center for Theoretical Physics, 34100 Trieste, Italy}

\abstract{\textBlue We present a global analysis of all the available
solar neutrino data treating consistently the $^8$B and $hep$ neutrino
fluxes as free parameters. The analysis reveals at $99.7$\% C.L. eight
currently-allowed discrete regions in two-neutrino oscillation space,
five regions corresponding to active neutrinos and three corresponding
to sterile neutrinos. Most of the allowed oscillation solutions are
robust with respect to changes in the analysis procedures, but the
traditional vacuum solution is fragile. The globally-permitted range
of the $^8$B neutrino flux, $0.45$ to $1.95$ in units of the BP2000
flux, is comparable to the $3\sigma$ range allowed by the standard
solar model. We discuss the implications for SNO of a low mass,
$\Delta m^2 \sim 6 \times 10^{-12}{\rm ~ eV^2}$, vacuum oscillation
solution, previously found by Raghavan, and by Krastev and Petcov, but
absent in recent analyses that included Super-Kamiokande data.  For
the SNO experiment, we present refined predictions for the
charged-current rate and the ratio of the neutral-current rate to
charged-current rate. The predicted charged-current rate can be
clearly distinguished from the no-oscillation rate only for the LMA
solution. The predicted ratio of the neutral-current rate to
charged-current rate is distinguishable from the no-oscillation ratio
for the LMA, SMA, LOW, and VAC solutions for active neutrinos.  }

\begin{document}
\textBlack
\input psfig
\def\lsim{\mathrel{\rlap{\lower4pt\hbox{\hskip1pt$\sim$}}
    \raise1pt\hbox{$<$}}}         %less than or approx. symbol
\def\gsim{\mathrel{\rlap{\lower4pt\hbox{\hskip1pt$\sim$}}
    \raise1pt\hbox{$>$}}}         %greater than or approx. symbol
\renewcommand{\topfraction}{.95}
\renewcommand{\bottomfraction}{.95}
\renewcommand{\textfraction}{0}
\renewcommand{\floatpagefraction}{1}
\hyphenation{Super-Kamiokande}

\section{Introduction}
\label{sec:introduction}
From the inception of the subject, solar neutrino research has been
motivated by two apparently conflicting goals: 1) to test the theory of
nuclear fusion reactions in stars; and 2) to determine neutrino
characteristics. In the approximately four decades since its
inception, the subject has been dramatically transformed.  In the
first paper reporting an experimental result~\cite{davis68}, the
measurement was compared only with the then existing standard solar
model~\cite{bahcall68}. In the ensuing decades, the emphasis gradually
shifted to particle physics as enormous progress was made both
experimentally and theoretically. New experiments were reported
\footnote{The total rates in the Homestake (chlorine), Kamiokande,
SAGE, GALLEX + GNO, and Super-Kamiokande experiments cannot be fit
well without some form of new physics even if the solar neutrino
fluxes are allowed to be free parameters. Allowing the $p-p$, $^7$Be,
$^8$B, $^{13}$N, and $^{15}$O fluxes to be free parameters, the
minimum $\chi^2$ is obtained for zero fluxes of $^7$Be $= ^{13}$N $=
^{15}$O $= 0.0$ and even this unphysical solution is acceptable only
at the $99.6$\% C.L. This result has been stable for many years as
experimental results have been refined.}, including the results of
Kamiokande~\cite{kamiokande}, SAGE~\cite{sage}, GALLEX~\cite{gallex},
Super-Kamiokande~\cite{superk}, GNO~\cite{gno}, refined results of the
chlorine experiment~\cite{chlorine}, and (in the near future) there
will be results from SNO~\cite{sno}, BOREXINO~\cite{borexino},
KamLAND~\cite{kamland} and ICARUS~\cite{icarus}. In parallel
activities, the theories of vacuum~\cite{vac} and matter-induced
(MSW)~\cite{msw} neutrino oscillations were developed and explored and
the solar models were refined~\cite{neutrinoastrophysics} and verified
by helioseismology~\cite{bp98}.

In the last decade or so, it has become customary to blur the
distinction between the two goals of solar neutrino research,
measuring neutrino properties and using neutrinos to learn about
stars. The results of all the experiments are combined in a
statistical analysis from which the allowed ranges of neutrino masses
and mixing angles are extracted, including among the input data the
calculated standard solar model neutrino fluxes and their associated
uncertainties.  

In the present paper, we take a modest step toward separating the two
subjects, neutrino physics and neutrino astronomy, of solar neutrino
research. We allow the important $^8$B neutrino flux, and the much
less important $hep$ flux, to be free parameters and perform a
systematic global analysis~\cite{bks98,flm98} of all the available
solar neutrino data (for early work allowing the $^8$B neutrino flux
to vary freely, see ref.~\cite{ks94} and for related work, see
refs.~\cite{bks98,concha99,GFM,concha2000}). We
extract from the analysis the allowed ranges of the $^8$B and $hep$
fluxes as well as the neutrino parameters, $\Delta m^2$ and $\tan^2
\theta$. We continue, following what is currently common practice, to
constrain the other solar neutrino fluxes with the aid of the
calculated fluxes and uncertainties given by the BP2000 standard solar
model~\cite{bp2000}.

We emphasize the robustness of most of the allowed regions, and the
fragility of some regions, to small changes in the data analysis. We
illustrate the effects of changes in the analysis by performing the
global analysis of all the data in different ways. In particular, we
demonstrate the effects of the common practices (of which we have also
been guilty) of treating the $^8$B absolute flux differently between
the measured rates and the measured spectral data and the effects of
double counting of the SuperKamiokande rate.

We have also carried out solutions in which the $^7$Be, $^8$B, and
$hep$ fluxes are all allowed to vary without taking into account the
solar model predictions, but in this case the range of solutions is
too large at present to be useful to discuss. The situation will
presumably change when data from the SNO, KamLAND, BOREXINO and ICARUS
experiments are available.  In a work in preparation, we will report
on the implications of the global solutions found here for $^7$Be
experiments like BOREXINO.

In Section~\ref{sec:global}, we present the global solutions for both
active and sterile neutrinos when the $^8$B and $hep$ fluxes are
treated as free parameters. Section~\ref{sec:variations} shows that
some solutions (LMA, SMA, and LOW) are robust with respect to changes
in the analysis constraints while other solutions (vacuum solutions)
are more fragile.  We discuss in Section~\ref{sec:justso} the
characteristics of the Just So$^2$ solution and in
Section~\ref{sec:sno} we present the predictions of the
currently-allowed oscillation solutions for the measurements with SNO
of the charged-current rate and the (charged-current rate)
/(neutral-current rate) ratio. We summarize and discuss our main results in
Section~\ref{sec:discussion}.

\FIGURE[!ht]{
\caption{Global solutions, free $^8$B and $hep$ fluxes.  (a)\ Active
neutrinos. (b) \ Sterile neutrinos. The input data include the total
rates measured in the Homestake, SAGE, and GALLEX + GNO experiments
and the electron recoil energy spectrum measured by Super-Kamiokande
during the day and also the spectrum measured at night.  The best-fit
points are marked by dark circles; the allowed regions are shown at
$90$\%, $95$\%, $99$\%, and $99.73$\% C.L. .
\label{fig:global}
}
\centerline{\raise2.8in\hbox{(a)}\psfig{figure=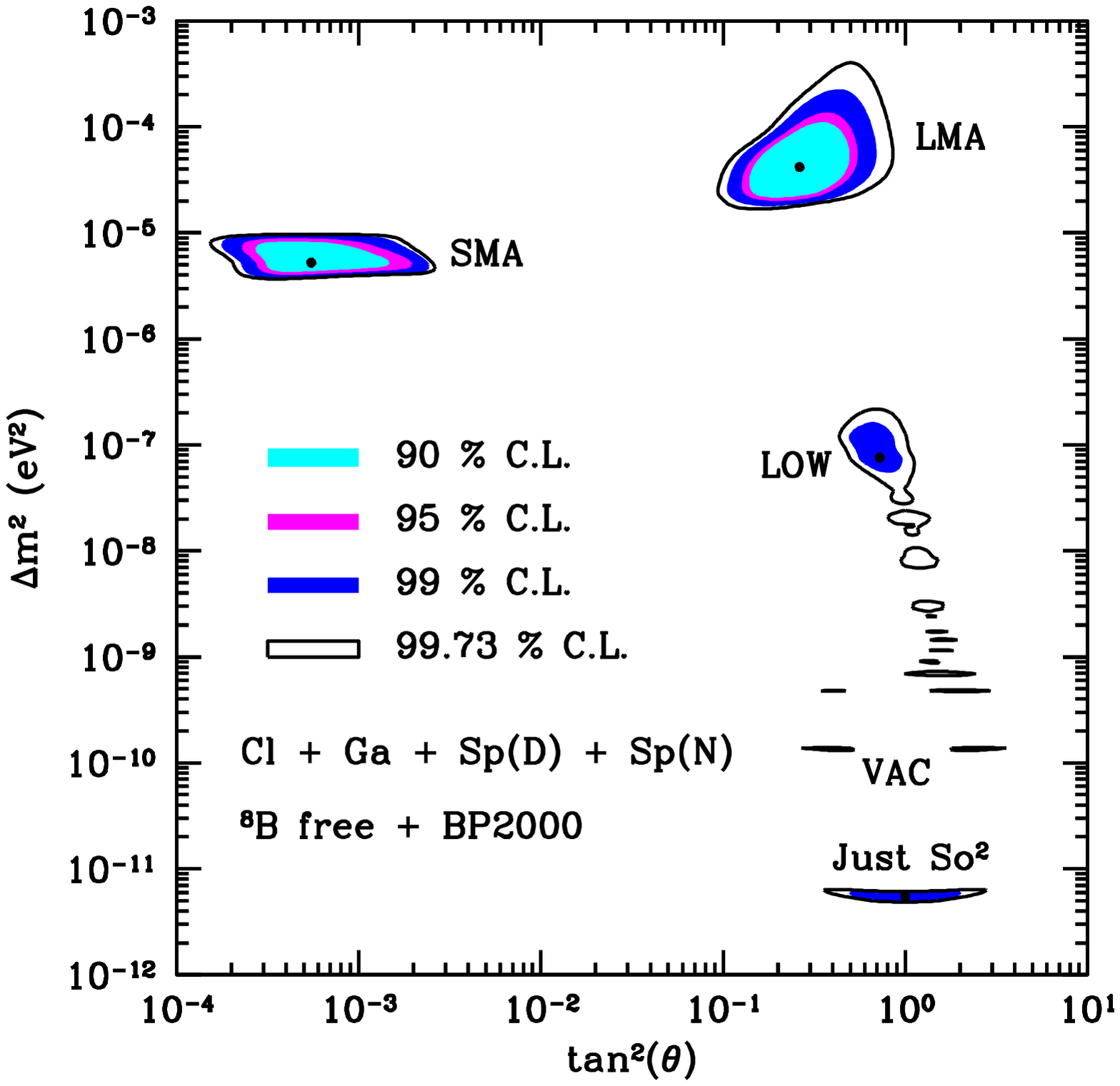,width=3.54in}}
\centerline{\raise2.8in\hbox{(b)}\psfig{figure=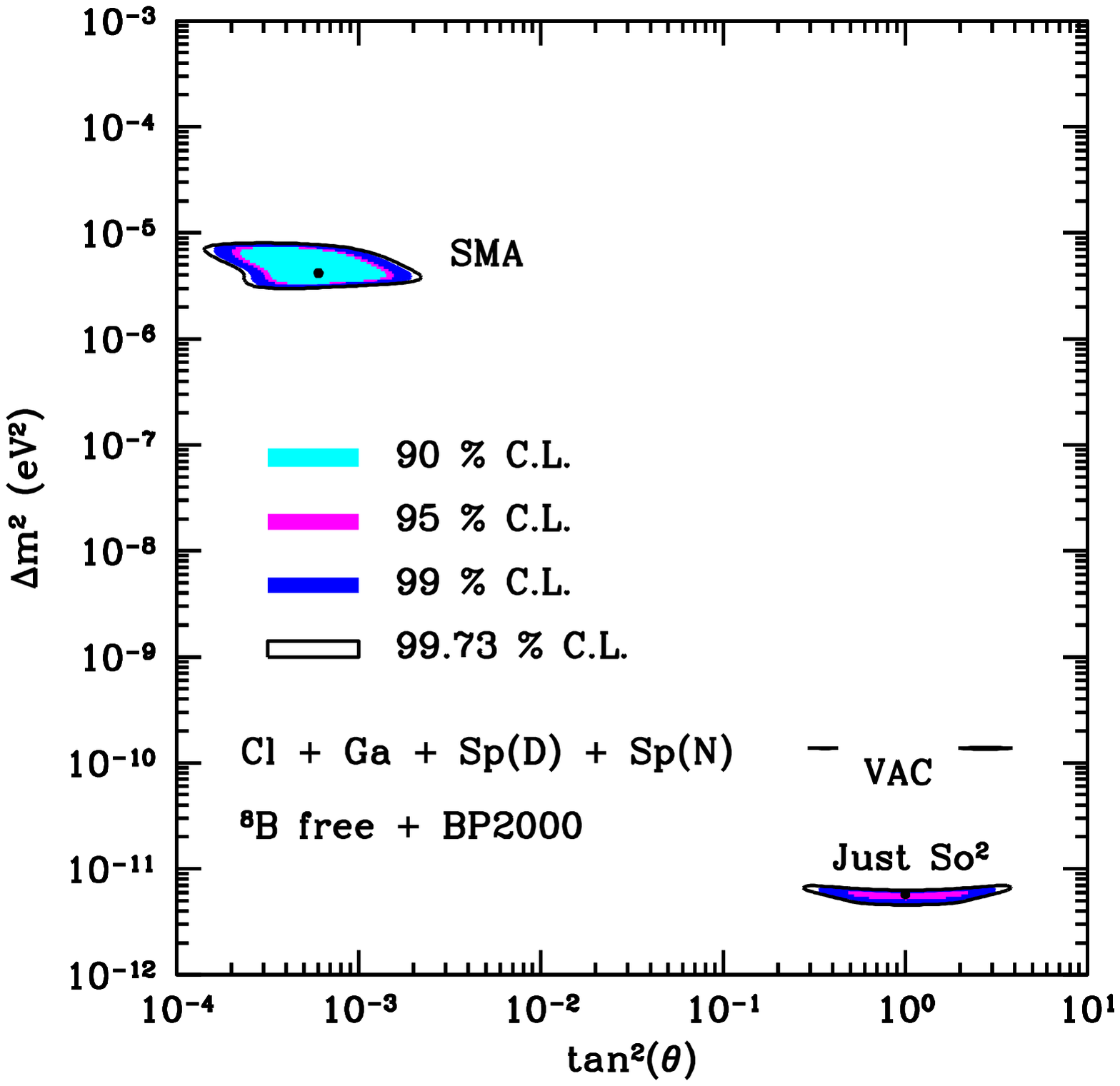,width=3.54in}}
}

\section{Global solutions}
\label{sec:global}

We summarize our main results on the global two-neutrino oscillation
solutions in Section~\ref{subsec:results} and describe in
Section~\ref{subsec:calculations}, which is intended for aficionados
only, our calculational procedures.

\subsection{Results}
\label{subsec:results}

Figure~\ref{fig:global} shows the globally allowed solutions for both
active, Figure~\ref{fig:global}a, and sterile neutrinos,
Figure~\ref{fig:global}b. The results are presented at four different
confidence levels ranging from $90$\% to $99.73$\% (corresponding to
$3\sigma$). There are five isolated regions of allowed solutions for
active neutrinos (LMA, SMA, LOW, VAC, and Just So$^2$) and three separate
solutions for sterile neutrinos (SMA, VAC, and Just So$^2$).

The allowed oscillation regions are shown for a global solution with
the $^8$B neutrino flux treated as a free parameter in all of the
analysis.  The allowed regions at different confidence levels are
presented for neutrino oscillation models that fit the total rates
measured in the chlorine~\cite{chlorine} and
gallium~\cite{sage,gallex,gno} solar neutrinos experiments, as well as
the electron recoil energy spectrum measured by the Super-Kamiokande
collaboration~\cite{superk} during the day and the spectrum measured
at night.  We treat as one experiment the combined results of the
GALLEX and GNO measurements and consider the SAGE results to be an
independent experiment.  We do not include in this analysis the
Super-Kamiokande total rate, since to a large extent the total rate is
represented by the flux in each of the spectral energy bins. However,
since many groups analyzing solar neutrino data include both the total
Super-Kamiokande rate and the spectral data, we perform the analysis
in this way in the following section, Section~\ref{sec:variations}.

The best-fit points in each region are shown as black dots. The
measurements and errors are taken from the publications of the
experimental groups.  We use in this paper solar neutrino data that
appeared in papers published before February 1, 2001 or in Neutrino
2000. The theoretical errors on all the other fluxes are taken from
the BP2000 solar model~\cite{bp2000}. The Super-Kamiokande measurement
for the $^8$B neutrino flux is $\phi(^8B) = (2.40 \pm 0.03
^{+0.08}_{-0.07}) \times 10^6 {\rm cm^{-2}\, s^{-1}}$.

Matter effects are significant for all of the allowed islands of
solution space between  $10^{-9} {\rm eV^2} \leq \Delta m^2 \leq 3\times
10^{-7} {\rm eV^2}$. We call this collection of islands the LOW
solution. In some ways of analyzing the data, all of the LOW islands
are surrounded by a single $3\sigma$ contour.

Following Fogli, Lisi, and   Montanino~\cite{lisitan} and 
de Gouvea, Friedland, and Murayama~\cite{GFM}, we have given our
results in terms of $\tan^2 \theta$ rather than $\sin^2 2\theta$ in
order to include solutions with mixing angles greater than $\pi/4$
(the so-called `dark side').  The general procedure that we have used
in deriving the allowed regions is described in ref.~\cite{bks98}; see
Section~\ref{subsec:calculations} for some details.

\TABLE{
\centering 
\caption{\label{tab:bestfits} {\bf Best-fit global oscillation
parameters.}  The oscillation solutions are
obtained by varying the $^8$B and $hep$ fluxes as free parameters in a
consistent way: simultaneously in the rates and in the night and day
spectrum fits.  The first five rows refer to active neutrinos (see
Figure~\ref{fig:global}a) and the last three rows refer to sterile
neutrinos (see Figure~\ref{fig:global}b).  The differences of the
squared masses are given in ${\rm eV^2}$. The number of degrees of
freedom is 35 [36(spectrum) + 3(rates) $-$4(parameters: $\Delta {\rm
m}^2$, $\theta$, and the $^8$B and $hep$ fluxes)].}
\begin{tabular}{lcccc} 
\noalign{\bigskip}
\hline 
\noalign{\smallskip}
Solution&$\Delta m^2$&$\tan^2(\theta)$& $\chi^2_{\rm min}$ &g.o.f. \\
\noalign{\smallskip}
\hline
\noalign{\smallskip}
LMA& $4.2\times10^{-5}$  &$2.6\times10^{-1}$ & 29.0 &$75$\% \\
SMA& $5.2\times10^{-6}$  &$5.5\times10^{-4}$ & 31.1 &$66$\%\\
 LOW& $7.6\times10^{-8}$  &$7.2\times10^{-1}$ & 36.0 &$42$\%\\ 
 Just So$^2$ & $5.5\times10^{-12}$  &\hskip -6pt$1.0\times10^{0} $ & 36.1 &$42$\%\\  
 VAC& $1.4\times10^{-10}$ &$3.8\times10^{-1}$ & 37.5 &$36$\%\\ 
 Sterile SMA & $4.2\times10^{-6}$ & $6.0\times10^{-4}$ & 32.5 & $59$\%\\
 Sterile Just So$^2$ & $5.5\times10^{-12}$  &\hskip -6pt$1.0\times10^{0} $ & 36.5 &$40$\%\\   
 Sterile VAC & $1.4\times10^{-10}$ & $3.6\times10^{-1}$ & 41.4 & $21$\%\\  
\noalign{\smallskip}
\hline
\end{tabular}
}

\begin{figure}
\centerline{\psfig{figure=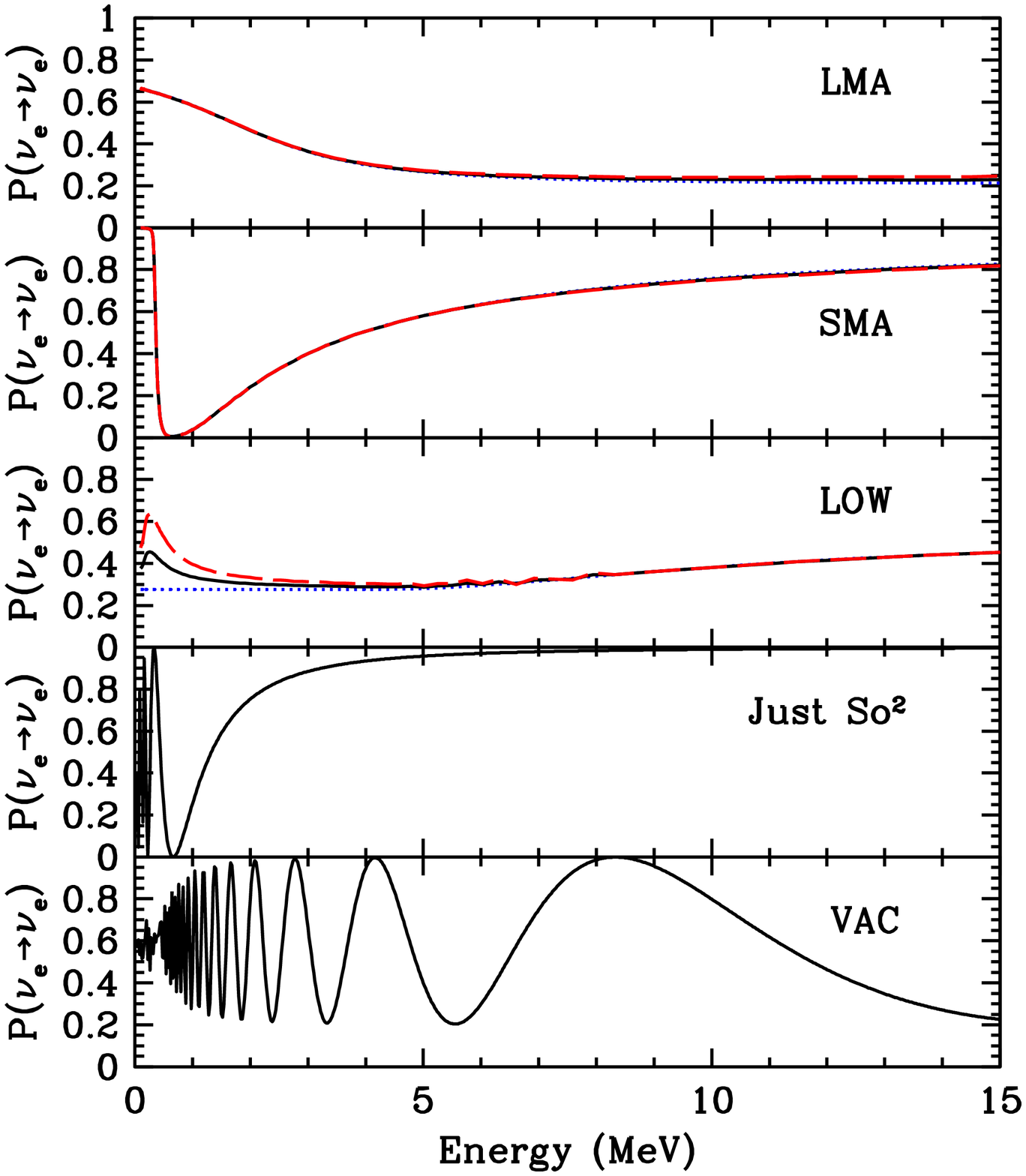,width=4.75in}}
\caption[]{Survival probabilities.  The figure presents the
yearly-averaged, best-fit survival probabilities for an electron
neutrino that is created in the sun to remain an electron neutrino
upon arrival at the earth. The survival probabilities for the sterile
solutions, SMA, Just So$^2$, and SMA, are very similar to their
counterparts for active neutrinos and are not plotted here.  The full
line refers to the average survival probabilities computed taking into
account regeneration in the earth and the dotted line refers to
calculations for the daytime that do not include regeneration.  The
dashed line includes regeneration at night.  There are only slight
differences between the computed regeneration probabilities for the
detectors located at the positions of Super-Kamiokande, SNO and the
Gran Sasso Underground Laboratory (see ref.~\cite{BK97}).
\label{fig:survival5}}
\end{figure}

Table~\ref{tab:bestfits} summarizes the properties of the best-fit
points of each allowed region: $\Delta m^2$, $\tan^2\theta$, and
goodness of fit, g.o.f. = 1 - C.L.  All of the solutions listed in
Table~\ref{tab:bestfits} and shown in Figure~\ref{fig:global} are
allowed at a comfortable confidence level. The LMA and the SMA
solutions are slightly preferred.

The global minimum, $\chi^2_{\rm min}/{\rm d.o.f.} =
29.0/35 = 0.83$, found here (see Table~\ref{tab:bestfits}) is somewhat
smaller than would be expected for data that have a true $\chi^2$
distribution with correctly estimated errors. The principal reason
that the $\chi^2$ is somewhat small is that the Super-Kamiokande day
and night recoil energy spectra are very well fit by undistorted $^8$B
and $hep$ energy spectra, $\chi^2_{\rm min} = 29.0$ for $34$ d.o.f.,
with $\phi(^8B) = 0.46 \phi(^8B)_{\rm BP2000}$ and $\phi(hep) = 1.0
\phi(hep)_{\rm BP2000}$ and C.L. of $29$\% .

Figure~\ref{fig:survival5} shows the computed survival probabilities
for electron-type neutrinos as a function of energy for the day (no
regeneration in the earth), the night (with regeneration), and the
annual average.  The probabilities were calculated for the best-fit
parameters listed in Table~\ref{tab:bestfits}.  The two most striking
aspects of this figure are the smallness of the day-night difference
(clearly visible in the figure only for the LOW solution at energies
below $1$ MeV) and the relative flatness (except for the fragile VAC
solution, see Section~\ref{sec:variations}) of the survival
probabilities at higher energies.

Most previous global analyses (including some analyses that we have
published) that took account of Super-Kamiokande data on the recoil
eneregy spectrum and the day-night effect have treated the $^8$B
absolute flux differently in fitting the spectral data and in fitting
the total rate.  In previous analyses, the $^8$B neutrino flux was
treated as a free parameter in fitting the Super-Kamiokande spectral
data but was treated as an input parameter, constrained by the
calculated standard solar model uncertainties, in fitting the data for
the measured total rates of the chlorine, gallium, and electron
scattering experiments. This lack of consistency was not present when
only rate data were fitted.

\TABLE{
\centering
\caption{\label{tab:fluxranges}{\bf Ranges of allowed fluxes}.  The
table lists the minimum (min) and maximum (max) values that are
allowed at $9 9.73$\% C.L.  for the $^8{\rm B}$ and $hep$ fluxes as well
as the best fit (bf) values within each of the allowed regions.  The
$^8$B and $hep$ fluxes were allowed to vary freely and
consistently; the other neutrino fluxes are constrained by the errors
given in the BP2000 solar model predictions. The first five rows refer
to active neutrinos (see Figure~\ref{fig:global}a) and the last three
rows refer to sterile neutrinos (see Figure~\ref{fig:global}b).  The
best-fit global solutions are shown as black dots in
Figure~\ref{fig:global}a and Figure~\ref{fig:global}b.}
\begin{tabular}{lcccccc}
\noalign{\bigskip}
\hline 
\noalign{\smallskip}
Solution &  $^8$B& $^8$B& $^8$B& $hep$
& $hep$&$hep$\\
&(bf)&(min)&(max)&(bf)&(min)&(max)\\
\noalign{\smallskip}
\hline
\noalign{\smallskip}
\noalign{\smallskip}
\hline
\noalign{\smallskip}
LMA     &  1.31   &  0.78     &  1.95    &  0.5  & 0.0 & 8.5 \\  
  
SMA     &  0.61   &  0.50     &  1.42    &  1.0  & 0.0 & 5.5 \\

LOW     &  0.87   &  0.74     &  1.08    &  0.75 & 0.0 & 3.5 \\
 
Just So$^2$  &  0.47   & 0.45      &  0.48      &  0.5 & 0.0 & 2.0 \\

VAC     &  0.55   &  0.53     &  0.81    &  0.25 & 0.0 & 4.0 \\  

Sterile SMA  &  0.62   & 0.49  &  1.25  &  1.0 & 0.0 & 5.5 \\

Sterile Just So$^2 $  &  0.47   & 0.44  &  0.49  &  0.5 & 0.0 & 2.5 \\

Sterile VAC  &  0.57   & 0.54  &  0.60  &  1.0 & 0.0 & 12.0 \\
\noalign{\smallskip}
\hline
\end{tabular}
}

Table~\ref{tab:fluxranges} shows, for each allowed oscillation region,
the total range of the $^8$B and $hep$ fluxes permitted at $99.73$\%
C.L. .  The allowed regions were identified in a search in which
$\Delta m^2$, $\tan^2 \theta$, and the $^8$B and $hep$ fluxes were all varied
freely. The tabulated values represent the minimum and maximum values
of the $^8$B fluxes anywhere within the designated allowed regions
defined by the four  free parameters.

The fluxes given in Table~\ref{tab:fluxranges} are the total fluxes
created at the sun and can therefore be directly compared with the
predictions of the standard solar model. In terms of the best-estimate
$^8$B neutrino flux from the BP2000 model ($5.05 \times 10^8~{\rm cm^2
\, s^{-1}}$), the total currently allowed range of solutions is,
according to Table~\ref{tab:fluxranges},

\begin{equation}
0.44~\leq \phi(^8B)_{\rm \nu-analysis}/\phi(^8B)_{\rm BP2000} ~ \leq 1.95 .
\label{eq:b8range}
\end{equation}
The corresponding $3\sigma$ range allowed by the error analysis of the
standard solar model is

\begin{equation}
0.52~\leq \phi(^8B)/\phi(^8B)_{\rm BP2000} ~ \leq 1.6 .
\label{eq:b8bprange}
\end{equation}
The range allowed by the global analysis of neutrino experiments is
slightly larger than the estimated $3\sigma$ uncertainties in the
standard solar model $^8$B neutrino flux prediction. The largest
allowed value of the $^8$B flux corresponds to neutrino parameters
within the LMA allowed domain and the smallest allowed value is
realized within the Just So$^2$ solution. The allowed range of the
$hep$ flux is

\begin{equation}
0.0~\leq \phi(hep)_{\rm \nu-analysis}/\phi(hep)_{\rm BP2000} ~ \leq 12.0 .
\label{eq:heprange}
\end{equation}
Because the uncertainty in the nuclear fusion cross section for the
$hep$ reaction is large and difficult to quantify, no estimated error
is given for the $hep$ neutrino flux in the standard solar model.

\subsection{Calculational method}
\label{subsec:calculations}

We calculate the global $\chi^2(f_B)$ = $\chi^2_R(f_B)$ +
$\chi^2_{Sp}(f_B)$, where the subscripts ``$R$'' and ``$Sp$'' stand
for ``Rates'' and ``Spectrum'', for each $\Delta {\rm m}^2$ and
$\tan^2\theta$ on a $201 \times 500$ lattice using 50 points per
decade in both $\tan^2\theta$ and $\Delta {\rm m}^2$. The validity of
the $\chi^2$ approach in this context, and some results of alterantive
approaches, are discussed in refs.~\cite{garzelli,creminelli}. The parameter
$\Delta m^2$ varies from $10^{-12} {\rm eV}^2$ to $10^{-3} {\rm eV}^2$
and $\tan^2\theta$ varies from $10^{-4}$ to $10^{1}$. The $^8$B
neutrino flux is treated as a free parameter and at each step of the
minimization process is kept the same in both individual $\chi^2$'s
for the rates and for the spectrum. The $\chi^2_R$ for the rates is
calculated using the prescription given in \cite{FL95}, with updated
uncertainties for the astrophysical parameters taken from
BP2000~\cite{bp2000}.  We do not include uncertainties in the $^8$B
flux since we treat this flux as a free parameter.  

For the calculation of $\chi^2_{ Sp}$, we use the separate day and
night spectra measured by the Super-Kamiokande collaboration and
presented at Neutrino 2000~\cite{YS}. The statistical and systematic
errors in the spectrum data are included as explained in \cite{bks98}
with the simple but important refinement of including separately the
correlated and uncorrelated systematic errors in the off-diagonal and
diagonal elements of the covariance matrix. We use the undistorted
spectrum shape for $^8$B neutrinos that is given in ref.~\cite{shape}
(see also the very similar spectral shape of ref.~\cite{ortiz}).

After the global $\chi^2_{min}$ is determined, we draw the
C.L. contours in the plane $\tan^2\theta$ -- $\Delta {\rm m}^2$ by
connecting points with equal $\chi^2$ = $\chi^2_{min}$ +
$\Delta\chi^2$, where $\Delta\chi^2$ = 4.605, 5.99, 9.21, 11.83 for
90, 95, 99 and 99.73 \% C.L. for two degrees of freedom (the neutrino
parameters$\tan^2\theta$ and $\Delta {\rm m}^2$). 

For oscillations into an active neutrino,  the survival probabilities
for electron neutrinos produced in the Sun to
arrive in the detector as an electron neutrino are calculated using
the electron number density, $n_e$, in the BP2000 model~\cite{bp2000}.  For
oscillations into sterile neutrinos, we use the effective density
$n_{\rm sterile} = n_e - n_n/2$, where $n_n$ is the number density of
neutrons in the BP2000 model~\cite{bp2000}. We calculate numerically
the survival probabilities, using a hybrid algorithm  in which different
approaches are used for different values of the parameter $E/\Delta
{\rm m}^2$. 

For ${\rm E}/\Delta {\rm m}^2$ $<$ $3 \times 10^6$ ${\rm MeV/eV}^2$
and all angles (including $\theta > \pi/4$ \cite{Alex,GFM}), we use the
well known analytical prescription \cite{KP88} for calculating the
survival probability at the surface of the Sun using the exact
analytical solution for exponential density profiles
\cite{Kaneko,Toshev,STP}. The survival probability was
averaged over the relevant production region for each neutrino flux
(e. g., $^8$B or $^7{\rm Be}$) as given in the BP2000
model~\cite{bp2000}.

For all other cases ( ${\rm E}/\Delta
{\rm m}^2$ $>$ $3 \times 10^6$ ${\rm MeV/eV}^2$),  first the transition
probability $P_\odot(\nu_e \rightarrow \nu_1)$ of an electron neutrino
to the $\nu_1$ neutrino mass eigenstate at the surface of the Sun was
obtained numerically by solving the system of evolution equations in
the form given in \cite{MS}. The same system of equations was used to
calculate the transition probability $P_\oplus(\nu_1 \rightarrow
\nu_e)$ in the earth. The final survival probability was obtained
using the formula \cite{FLMP}:

\begin{eqnarray}
\lefteqn{P(\nu_e \rightarrow \nu_e) =} \nonumber\\
&&P_\odot(\nu_e \rightarrow \nu_1)
P_\oplus(\nu_1\rightarrow\nu_e) + ( 1 - P_\odot(\nu_e \rightarrow
\nu_1))(1 -P_\oplus(\nu_1\rightarrow\nu_e)) + \nonumber\\ 
\nonumber \\
&&2 \sqrt{P_\odot(\nu_e \rightarrow \nu_1)
P_\oplus(\nu_1\rightarrow\nu_e)( 1 - P_\odot(\nu_e \rightarrow
\nu_1))(1 -P_\oplus(\nu_1\rightarrow\nu_e))} \cos\phi, \nonumber \\
\end{eqnarray}
\noindent
where $P_\odot(\nu_e \rightarrow \nu_1)$ is the transition probability
that a $\nu_e$ in the solar interior becomes a $\nu_1$ mass eigenstate
at the solar surface, and where $P_\oplus(\nu_1\rightarrow\nu_e)$ is
the transition probability that a $\nu_1$ becomes a $\nu_e$ after
crossing the earth.  The quantity $\phi$ is the phase difference of
the amplitudes of the $\nu_e \rightarrow \nu_1 \rightarrow \nu_e$ and
$\nu_e \rightarrow \nu_2 \rightarrow \nu_e$ transitions; the phase
difference is acquired as the neutrino states propagate between the
center of the Sun and the detector on Earth.  The phase is calculated
numerically at each stage of the propagation of the neutrino.  In the
region of parameter space where $E/\Delta {\rm m}^2 > 3 \times 10^6
{\rm MeV}/{\rm eV}^2$, averaging over the production region is
unnecessary since the transitions take place far from the region of
production.

The Earth regeneration effect is relevant for a rather limited range
of $E/\Delta m^2$, which is: $10^5 {\rm ~MeV}/{\rm eV}^2 < E/\Delta m^2
< 10^8 {\rm ~MeV}/{\rm eV}^2$. In this region of parameter space we
use the numerical procedure described in detail in ref.~\cite{BK97}. We
calculate the transition probabilities along a number of trajectories
(we use 0.5 degree spacing between adjacent trajectories) and average
them for each detector by using accurately calculated weights
proportional to the time the sun spends at different angles during the
course of a year.

\begin{figure}[!htb]
\caption[]{Global solutions including Super-Kamiokande rate and with
free $^8$B and $hep$ fluxes.  (a)\ Active neutrinos. (b) \ Sterile
neutrinos. The input data include the total rates measured in the
Homestake, SAGE, GALLEX + GNO, and Super-Kamiokande experiments and the
electron recoil energy spectrum measured by Super-Kamiokande during
the day and also the spectrum measured at night.  The best-fit points
are marked by dark circles; the allowed regions are shown at $90$\%,
$95$\%, $99$\%, and $99.73$\% C.L. .
\label{fig:globalsk}
}
\centerline{\raise2.8in\hbox{(a)}\psfig{figure=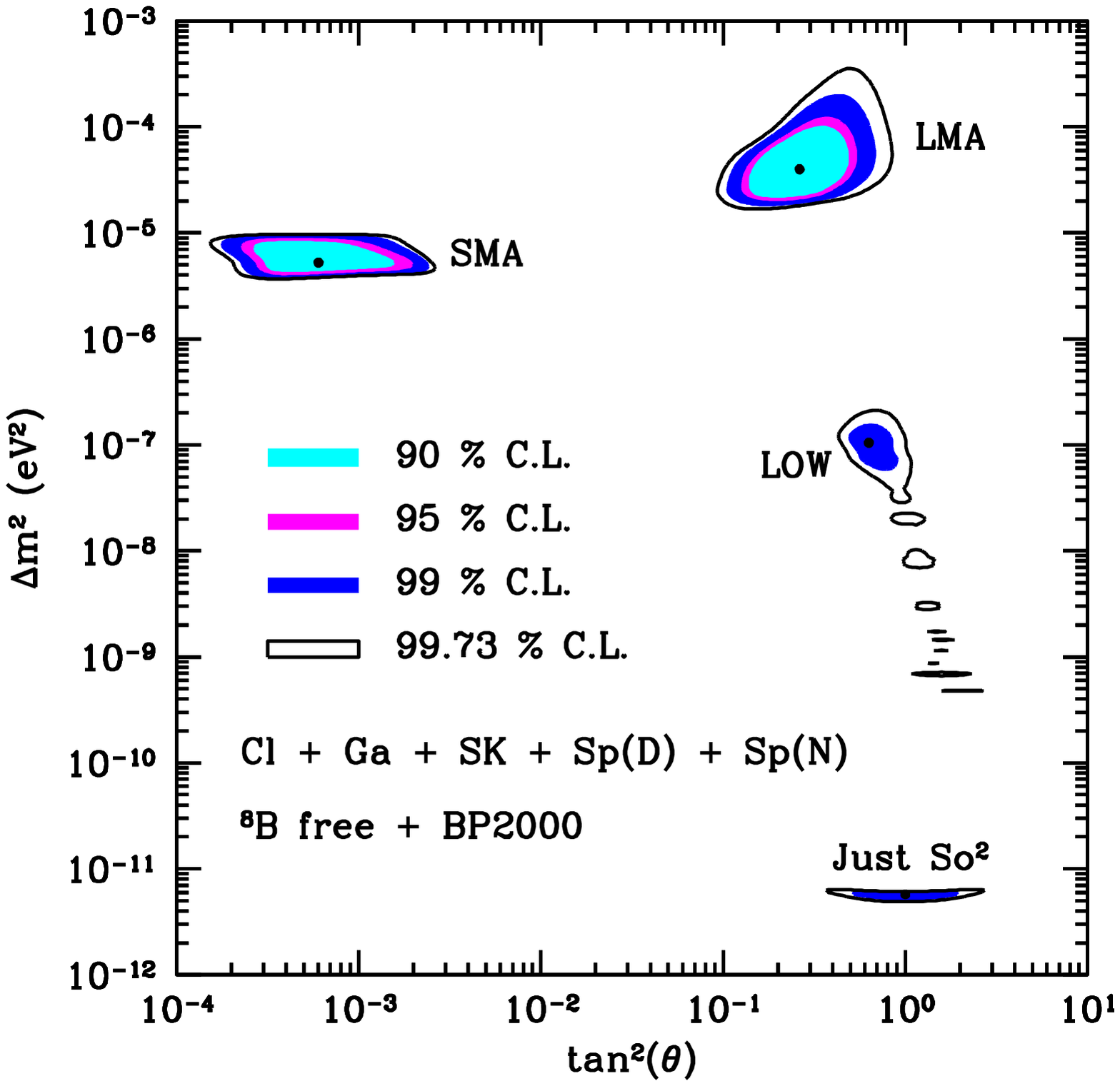,width=3.5in}}
\centerline{\raise2.8in\hbox{(b)}\psfig{figure=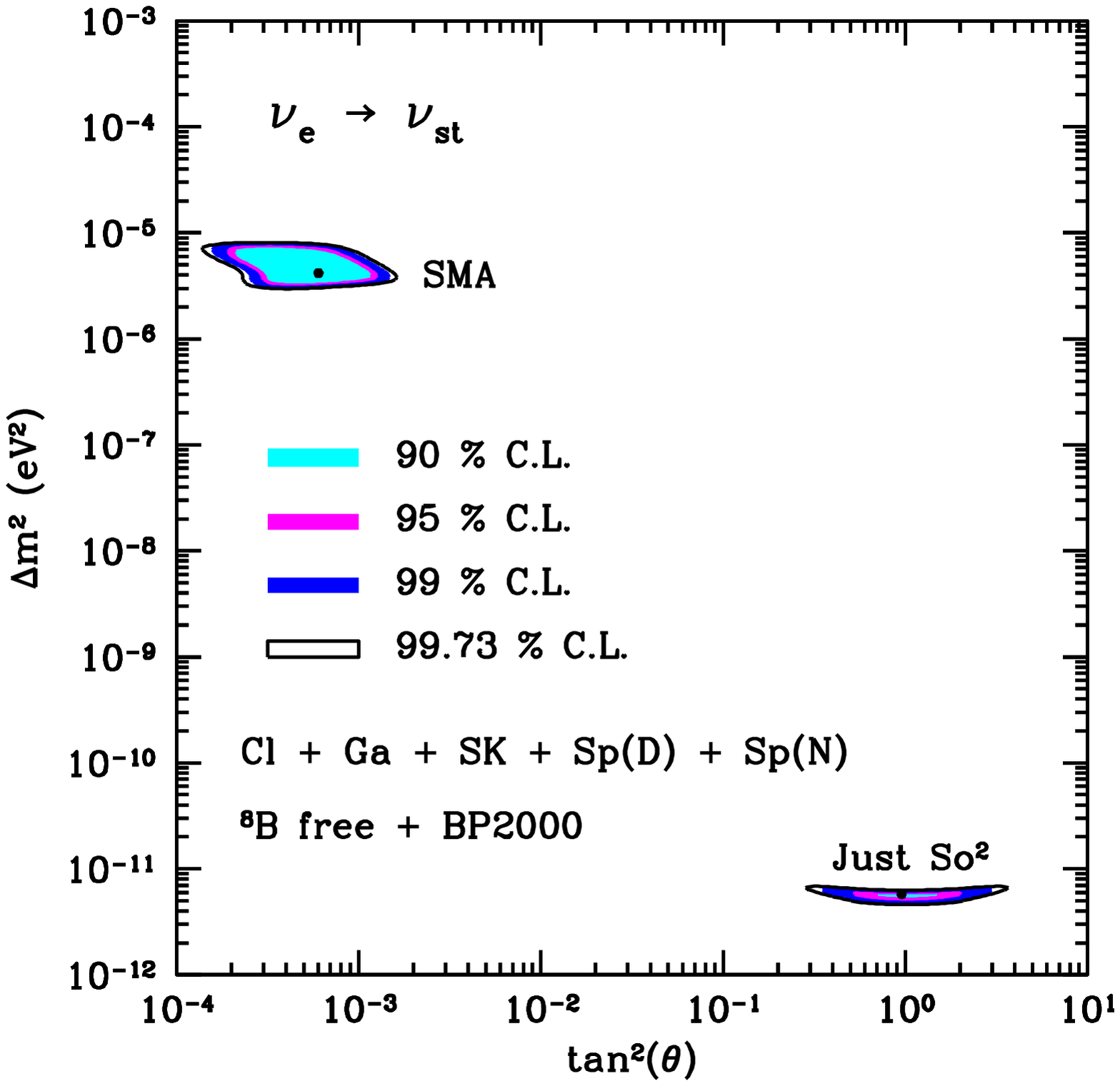,width=3.5in}}
\end{figure}

After the neutrinos leave the Sun and before they reach the detector, 
they oscillate in vacuum. The vacuum oscillations can be averaged over
energy analytically for all relevant $E/\Delta {\rm m}^2$, except for
oscillation lengths comparable to or larger than about 1 A.U. Seasonal
effects can be important for these longer oscillation lengths. The
analytical formula for an exponential density profile already includes
this averaging over vacuum oscillations and no additional averaging is
necessary when using this formula. In the region where the survival
probabilities are calculated numerically, the averaging is done by
propagating the neutrino state in vacuum over one oscillation length
and then taking the average of the periodic survival probability over
the same distance. Since the equations describing neutrino
oscillations in vacuum are exactly solvable, we use a simple
analytical expression for the average survival probability. In the
region $E/\Delta {\rm m}^2 > 5\times 10^{8} {\rm MeV}/{\rm eV}^2 $,
we include the oscillations in vacuum using the
one year averaged survival probabilities for which a convenient
analytical expression exists \cite{FFLM}.

\section{Variations on a theme}
\label{sec:variations}
In this section, we illustrate the extent to which the allowed
oscillation regions are robust or fragile by performing the global
analysis in different ways that have been used in the literature. 

We do not repeat here a misleading procedure that has sometimes been
used in the literature. It is incorrect to apply an exclusion region
at a fixed confidence level based upon the results of a particular
measured quantity (for example spectral data or day-night data) to an
allowed region based upon consistency with other measured quantities
(e. g., total rates). All of the measured quantities should be
analyzed together in a single global fit, which is the procedure
we follow in this paper.

Figure~\ref{fig:globalsk} presents the global solution for the case in
which the Super-Kamiokande total rate is included together with the
recoil electron energy spectrum. This double-counting procedure has
been adopted in many analyses in the literature, including analyses
that we have published.  It would be correct to use both the total
rate and the rates in each spectral bin if the total rate could be
determined in a way that was independent of the spectral
measurements. Since this is not the case~\cite{yochiro}, we have
chosen as our standard analysis in this paper the results shown in
Figure~\ref{fig:global} in which only the spectral data are used for
Super-Kamiokande.

Comparing Figure~\ref{fig:global} and Figure~\ref{fig:globalsk}, we
see that the six most probable allowed regions (LMA, SMA, LOW, and
Just So$^2$ for active neutrinos and SMA and Just So$^2$ for sterile
neutrinos, cf. Table~\ref{tab:bestfits}) are essentially unaffected by
whether or not one includes the Super-Kamiokande total rate in the
global analysis. The only qualitative change is that the least
probable solutions in Figure~\ref{fig:global}, the vacuum solutions at
$\Delta m^2 \sim 10^{-10}{\rm ~ eV^2}$, are absent if one includes the
Super-Kamiokande rate.

\FIGURE[!t]{
\caption{Influence of constraints on global solutions.  (a)\ Rates
only, $^8$B flux free. (b) \ $^8$B flux constrained by BP2000
uncertainty.  The calculations are the same as for
Figure~\ref{fig:globalsk}a except for one difference per panel.  For
Figure~\ref{fig:comparison}a, only total rates were considered and for
Figure~\ref{fig:comparison}b, the total $^8$B flux was constrained by
the BP2000 standard solar model uncertainty in calculating the
contribution of the rates to the total $\chi^2$ but was allowed to
vary to fit the spectrum. 
\label{fig:comparison}
}
\centerline{\raise2.8in\hbox{(a)}\psfig{figure=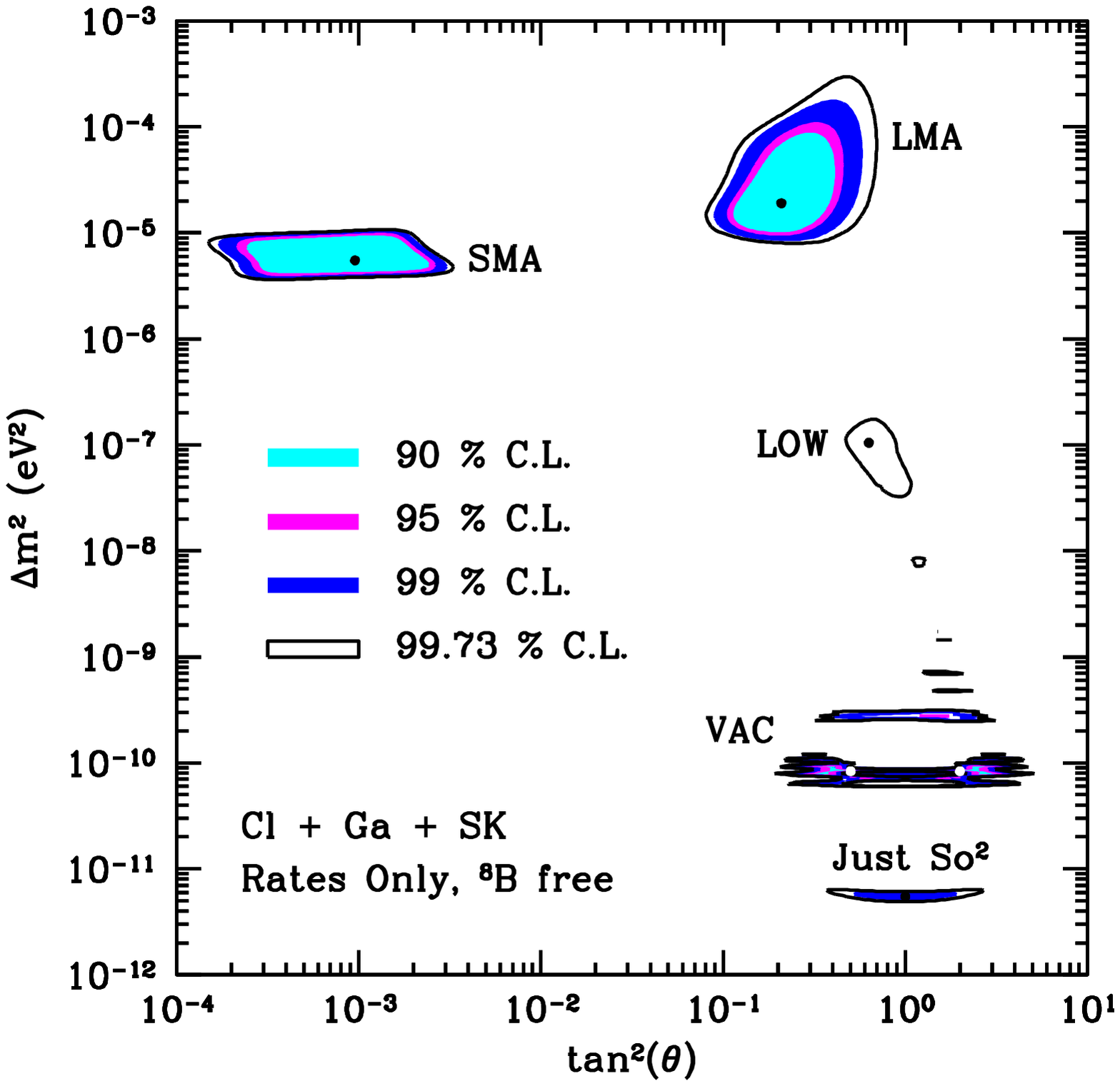,width=3.5in}}
\centerline{\raise2.8in\hbox{(b)}\psfig{figure=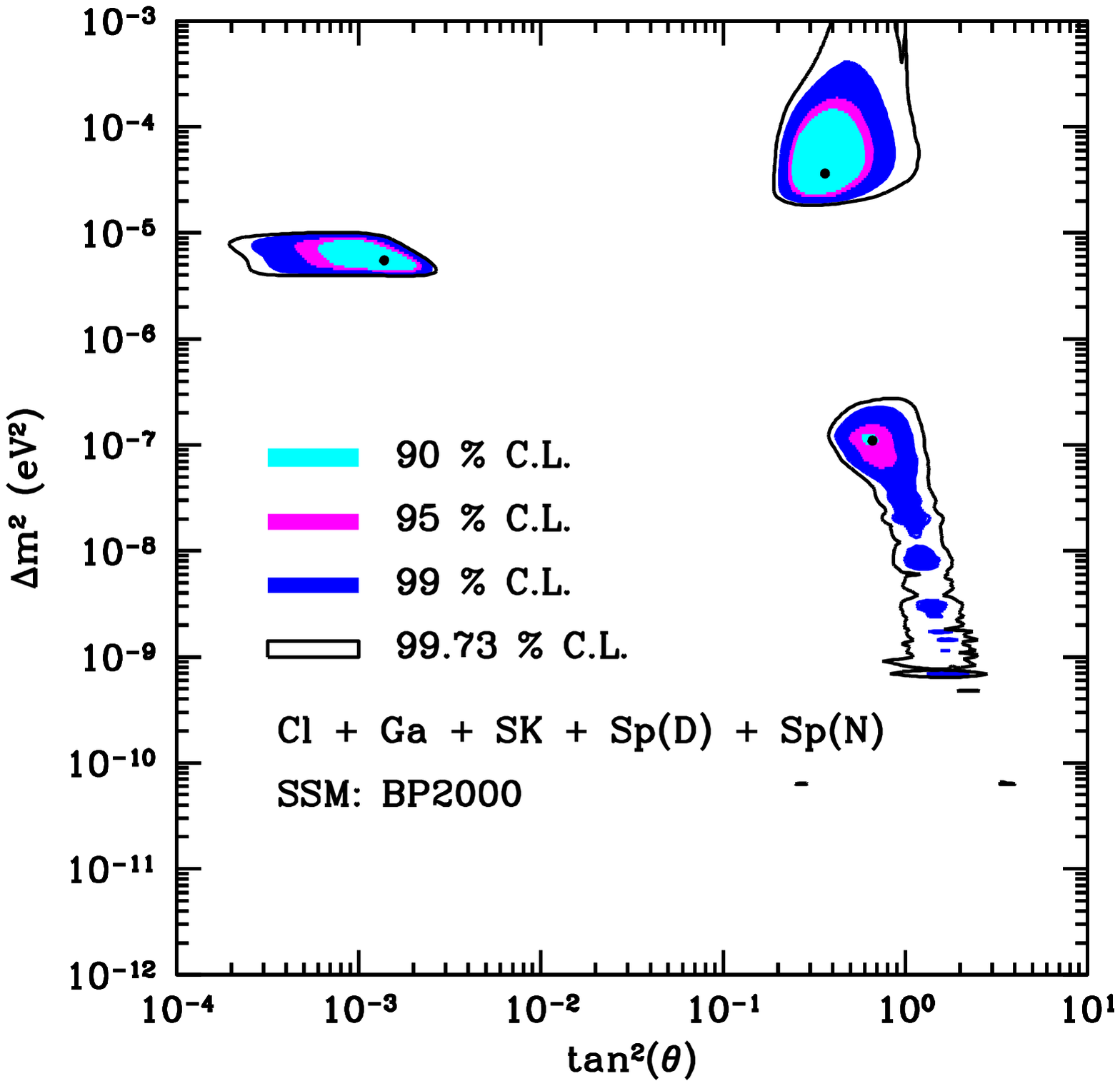,width=3.5in}}
}

Figure~\ref{fig:comparison} illustrates how two different constraints
affect the global solutions. The vacuum solutions with $\Delta m^2
\sim 10^{-10}~ {\rm eV^2}$ are prominent in
Figure~\ref{fig:comparison}a, which was constructed in the same way as
Figure~\ref{fig:globalsk}a except that for
Figure~\ref{fig:comparison}a only total rates, no spectral data, were
considered. Comparing Figure~\ref{fig:globalsk}a and
Figure~\ref{fig:comparison}a, we see clearly that the spectral data
have removed the previously prominent vacuum solutions.  The symmetric
best-fit points of the vacuum solution are marked by open circles in
Figure~\ref{fig:comparison}a.  

The only difference between the calculations that led to to
Figure~\ref{fig:comparison}b and to Figure~\ref{fig:globalsk}a is that
for Figure~\ref{fig:globalsk}a the BP2000 uncertainty for the $^8$B
neutrino flux was included in evaluating the contribution to the total
$\chi^2$ of the individual rates.  The imposition of the SSM flux
constraint decreases somewhat the goodness of fit of the
solutions. The best-fit points are shifted and the allowed regions are
distorted.  Also, including the standard solar model constraints
causes the LMA allowed region to overlap maximal mixing and to extend
to larger $\Delta m^2$ values (up to the region excluded by reactor
experiments, see the upper right corner of Figure~\ref{fig:comparison}b).

Most dramatically, constraining the $^8$B neutrino flux while comparing
the predictions to the total rates, eliminates the Just So$^2$
solution from Figure~\ref{fig:comparison}b.

We conclude that the LMA, SMA, and LOW solutions for active neutrinos,
and the SMA solution for sterile neutrinos, are all relatively
robust. They have been present since the first global analysis that
included Super-Kamiokande spectral and day-night data as well as the
total rates in the radiochemical experiments~\cite{bks98}. 

The vacuum solutions at $\Delta m^2 \sim 10^{-10}~ {\rm eV^2}$, on the
other hand, are relatively fragile. Whether or not the vacuum
solutions are allowed depends upon how much one emphases the
Super-Kamiokande data in the theoretical analysis.  The vacuum
solutions are present very prominently in the analysis if only the total
rates are considered (see Figure~\ref{fig:comparison}a and
ref.~\cite{bks98}), barely present if one includes the spectra data
but not the total Super-Kamiokande rate (see Figure~\ref{fig:global}),
and absent if one includes both the Super-Kamiokande rate and spectral
data (see Figure~\ref{fig:globalsk}).

The Just So$^2$ solutions, vacuum and sterile, are allowed if one
treats the $^8$B neutrino flux consistently as a free parameter in
fitting both the total rates and the Super-Kamiokande spectral data.

\section{Just So$^2$ Solution}
\label{sec:justso}

\FIGURE[!htb]{
\caption{Just So$^2$ vs BP2000. The survival probability for the
best-fit Just So$^2$ solution (dot-dashed lined) is shown versus 
the scaled neutrino fluxes (continuous lines) predicted by the
BP2000 solar model. The shapes of the continuous neutrino energy spectra are
correct but the fluxes have been scaled by constant values in order to
fit conveniently onto the same linear figure. The relative intensities
of the $^7$Be and $p-p$ lines are the same as in 
the BP2000 model.\label{fig:spectrum}}
\centerline{\psfig{figure=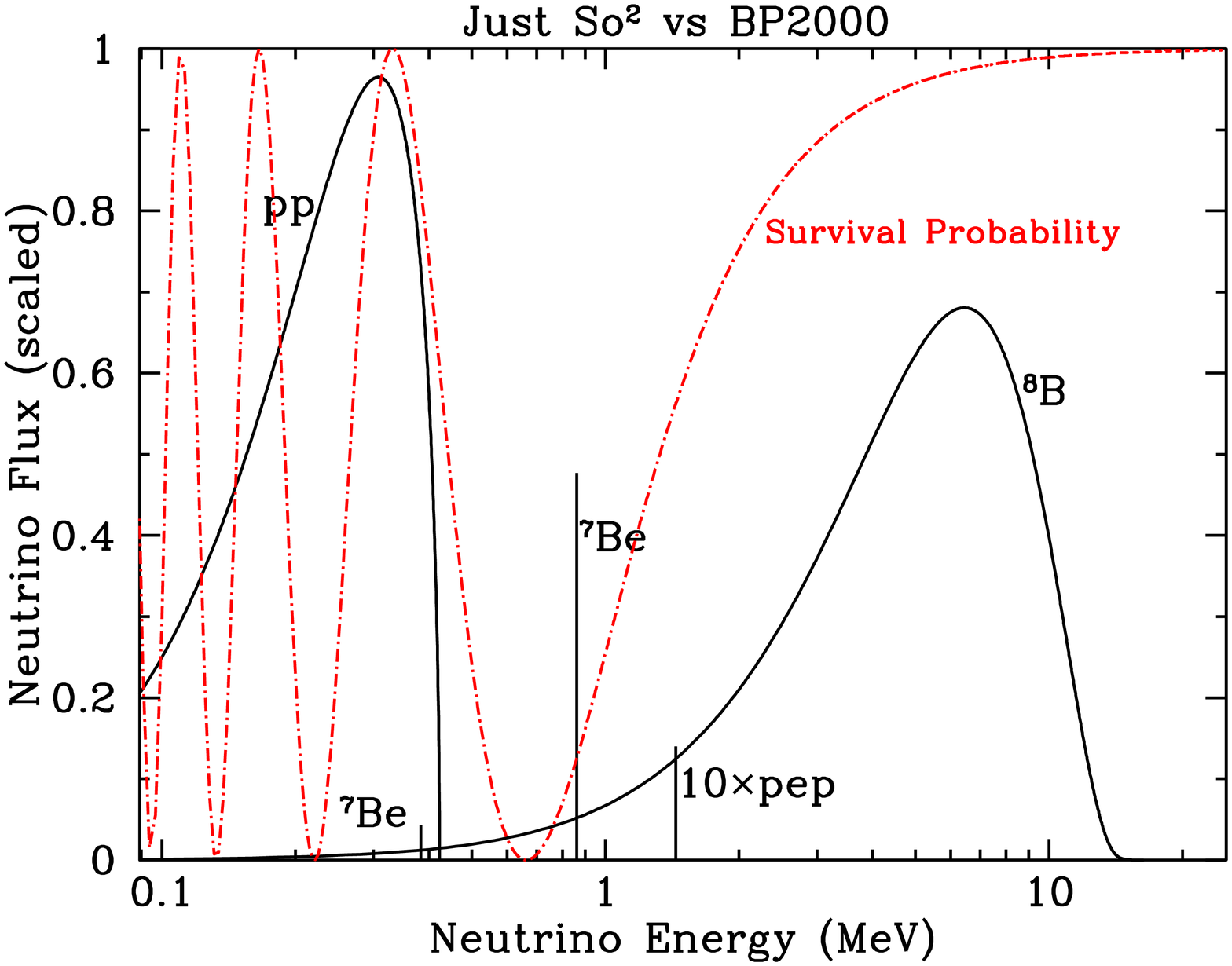,width=4in,angle=270}}
}

Figure~\ref{fig:global} contains two solutions, one for active
neutrinos and one for sterile neutrinos, that only appear if the $^8$B
flux is allowed to vary freely, namely, the solutions labeled ``Just
So$^2$.''  These solutions correspond to a best-fit mixing angle of
$\theta = \pi/4$ and a very small squared mass difference of $\Delta
m^2 ~\sim~ 6\times10^{-12}~{\rm eV^2}$
(cf. Table~\ref{tab:bestfits}). At the end of this section, we discuss
briefly the history of the Just So$^2$ solutions~\cite{raghavan,KP96}
and why they are not present in most analyses.

Figure~\ref{fig:spectrum} compares the Just So$^2$ survival
probability with the principal features of the solar neutrino
spectrum, namely, the two most important continuum fluxes ($p-p$ and
$^8$B) and the $^7$Be and $pep$ neutrino lines. The line fluxes are
expressed in units of $10^{10}~{\rm cm^{-2}\,s^{-1}}$. The continuum
fluxes have the correct energy dependence but are multiplied by
different constants, so that all the fluxes will fit conveniently onto
the same figure with a linear vertical scale.

The reason for using the name Just So$^2$ is apparent from
Figure~\ref{fig:spectrum}. The value of $\Delta m^2$ is just such that
the $^7$Be ($0.86$ MeV) $\nu_e$ survival probability is very small
($\sim 10$\%) and the $\nu_e$ survival probability at the peak ($0.31$
MeV) of the $p-p$ spectrum is very large ($\sim 87$\%).

\parbox[t]{4.5in}{

\TABLE[!t]{
\caption{\label{tab:justso} {\bf Just So$^2$ solution.}  
The table lists, for the best-fit Just So$^2$ solution, the contribution of
each flux to the chlorine and gallium experiments.}
\begin{tabular}{l@{\extracolsep{\fill}}ccc}
\noalign{\medskip}
\hline
\hline
\multicolumn{1}{c}{Neutrino}&Cl&Ga\\
\multicolumn{1}{c}{Source}& (SNU)&(SNU)\\
\noalign{\medskip}
\hline
\noalign{\medskip}
$p-p$&--&$55.7$\cr
$pep$& $0.1$&$1.6$\\
${\rm ^7Be}$& $0.1$&$5.1$\\
${\rm ^8B}$ & 2.7&5.6\\
${\rm ^{13}N}$& 0.02&0.7\\
${\rm ^{15}O}$& 0.16&2.2\\
\noalign{\smallskip}
\hline
\noalign{\smallskip}
Total&3.1&70.9\\
\noalign{\smallskip}
\hline
\noalign{\smallskip}
Observed&$2.56 \pm 0.21$&$74.7 \pm 5.1$\\
\end{tabular}
\label{tab:justsotable}
}}

Figure~\ref{fig:spectrum} provides an intuitive way of understanding
all the available solar neutrino experimental results. The lack of
spectral energy distortion measured by Super-Kamiokande above $5$ MeV is a
direct result of the smallness of the assumed $\Delta m^2$;
practically no oscillations occur above $5$ MeV. There is no predicted
measurable day-night effect because matter effects are all negligible
at such a small $\Delta m^2$. The SAGE and GALLEX plus GNO results are
accounted for by having the $^7$Be $\nu_e$ flux almost entirely absent
while the $p-p$ $\nu_e$ flux is hardly diminished.  The difference in
the ratio of the predicted standard rate to the measured rate in the
chlorine experiment (where it is a factor of three) and the
Super-Kamiokande experiment (where it is a factor of two) is explained
by the almost complete disappearance of the $^7$Be contribution to the
chlorine experiment.

Table~\ref{tab:justso} gives the contributions of the individual
neutrino fluxes to the chlorine and gallium experiments. The Just
So$^2$ solution does not provide an excellent fit to the chlorine
rate, but does provide very good fits to the rates of the gallium and
Super-Kamiokande experiments.  For Super-Kamiokande, the Just So$^2$
solution predicts a rate that is $0.461$ of the standard model rate,
in good agreement with the measured value~\cite{superk} of $0.475 \pm
0.016$.

The Just-so$^2$ solution is allowed for both active and sterile
neutrinos, with similar oscillation parameters and goodnes of fit.  In
general, the difference between active and sterile solutions is due to
the $\nu_{\mu}$ and $\nu_{\tau}$ that result from $\nu_e$
conversion. The $\nu_{\mu}$ and $\nu_{\tau}$ can contribute to $\nu -e
$ scattering in SuperKamiokande.  For the Just-So$^2$ solution, the
oscillation effect is practically absent at energies for which
SuperKamiokande is sensitive and therefore the $\nu_{\mu}$ and
$\nu_{\tau}$ fluxes do not contribute significantly even for active
neutrinos. This is the reason that for the experiments performed so
far (but not for BOREXINO), there is no appreciable difference between
the active and sterile cases for the Just So$^2$ solution.

Glashow and Krauss~\cite{gk} proposed the name of `Just So' neutrino
oscillations to describe vacuum oscillations for a neutrino mass
difference of $\Delta m^2 ~ = (50-130)\times10^{-12}~{\rm eV^2}$. The
mass of $\Delta m^2$ was chosen by Glashow and Krauss so as to greatly
reduce the $^8$B contribution to the chlorine experiment, assuming the
validity of the standard solar model.  For the Just So$^2$ solution
considered here, the $^8$B flux is assumed, when produced at the sun,
to already be significantly lower than predicted by the best standard
solar model.  The best-fit value of $\Delta m^2 ~\sim~
6\times10^{-12}~{\rm eV^2}$ suppresses strongly the contribution of
the $^7$Be neutrinos to the chlorine and gallium experiments, but
(unlike the Glashow-Krauss solution) does not affect the small $^8$B
flux assumed to be produced at the sun.

The Just So$^2$ solution was first found by Raghavan~\cite{raghavan}
and discovered independently and first analyzed in detail by Krastev
and Petcov~\cite{KP96}, who allowed the $^8$B flux to vary and
compared the results with the total rates measured in the chlorine,
Kamiokande, and gallium experiments (see also ref.~\cite{lp97}). No
spectral data or day-night effects were available when this analysis
was performed.  The reason that the Just So$^2$ solution was not found
in subsequent global solutions that included Super-Kamiokande spectral
data is that for Just So$^2$ the $^8$B flux is $3.3\sigma$ below the
the standard solar model~\cite{bp2000} flux. In many previous global
analyses, the $^8$B flux was allowed to vary in fitting the spectral
data but was constrained by the standard model uncertainties in
fitting the rate data~\cite{snoshow,fogli2000}.  The Just So$^2$
solution does appear in Figure~8 of our analysis~\cite{bks98} with a
free $^8$B flux of the total rates in the chlorine, gallium, and
Super-Kamiokande experiments, but was not found in the same work when
spectral and day-night data were included and the $^8$B flux was
constrained by the standard solar model uncertainty.

\section{Implications for the SNO experiment}
\label{sec:sno}

\FIGURE[!ht]{
\caption{Comparison of the CC~ SNO rate and the no oscillation
prediction.  The shaded area is the no oscillation prediction based
upon the measured Super-Kamiokande rate for $\nu-e$ scattering.  The
SNO CC ratios, [CC] = (to be measured)/(BP2000), are shown on the
vertical axes for different neutrino scenarios and two different total
electron energy thresholds, 5 MeV and 8 MeV. The error bars on the
neutrino oscillation results represent the range of values predicted
by the $99.73$\% CL allowed neutrino oscillation solutions displayed
in Fig.~\ref{fig:global}.\label{fig:cc}}
\centerline{\raise2.8in\hbox{(a)}\psfig{figure=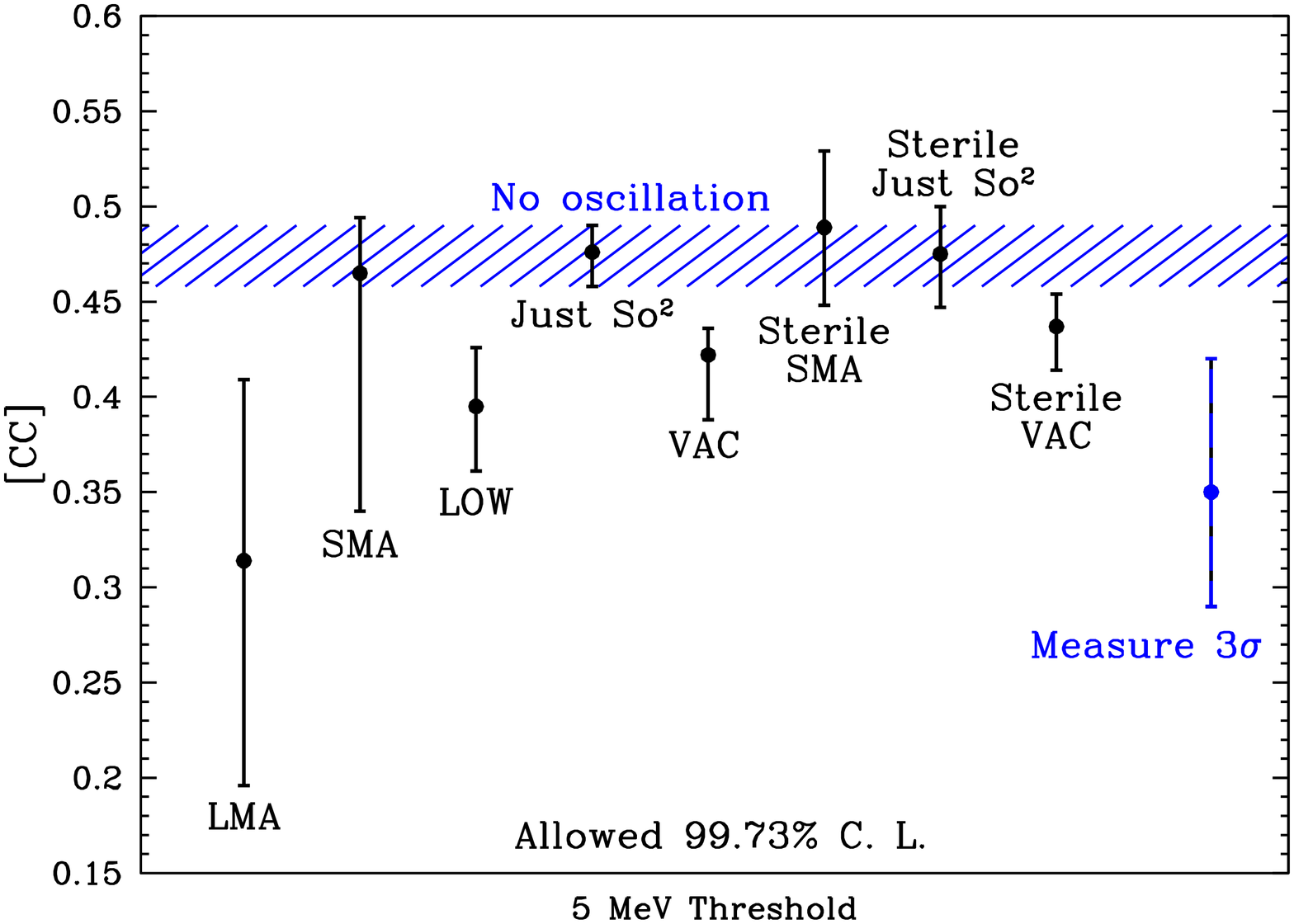,width=4in,angle=270}}
\centerline{\raise2.8in\hbox{(b)}\psfig{figure=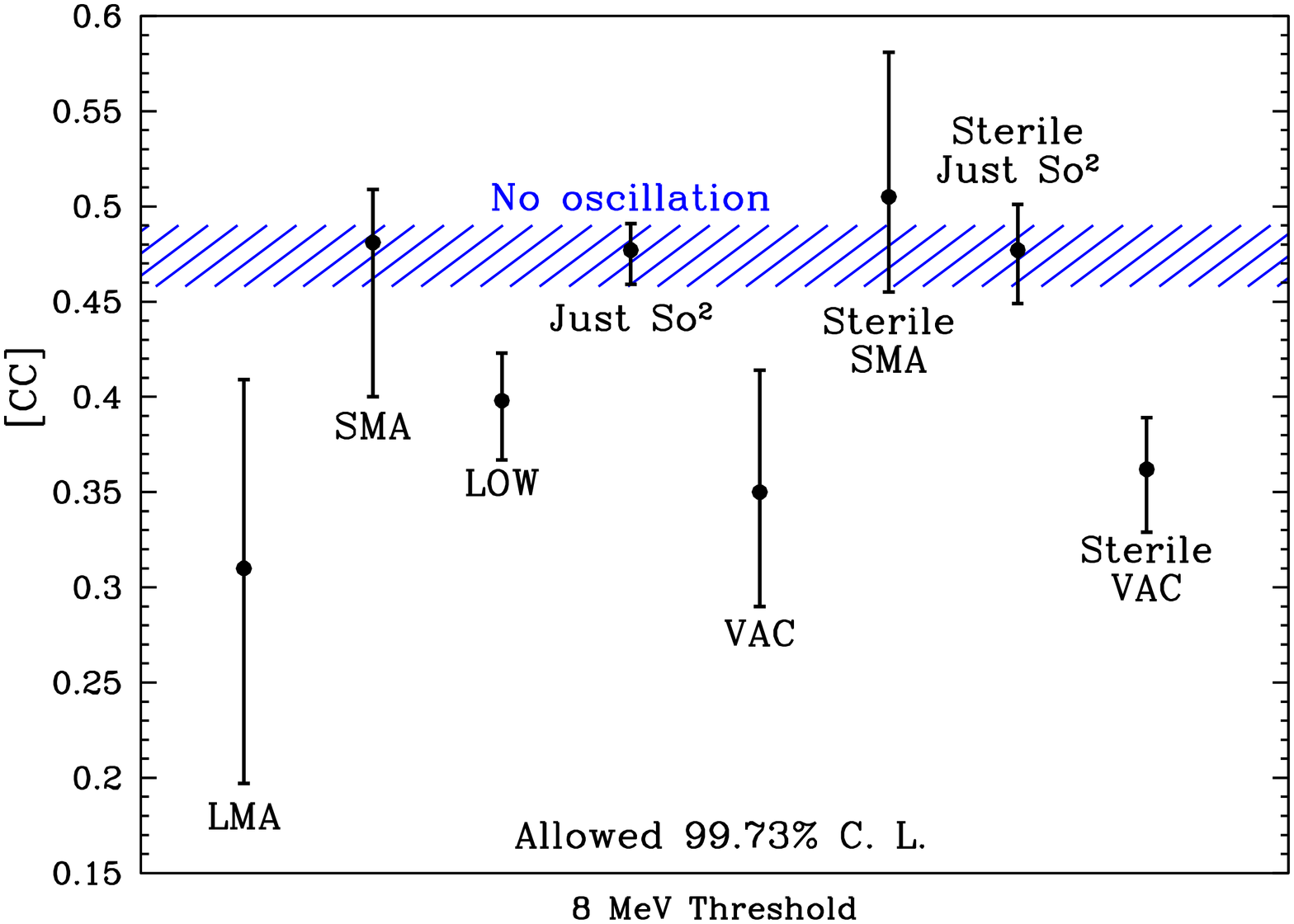,width=4in,angle=270}}
}

In this section, we first discuss the predictions for the
charged-current current rate in SNO and then discuss the predictions
for the ratio of the neutral-current rate to the charged-current
rate. We use the solutions that are allowed at $99.73$\% C. L. in the
global fit that is shown in Figure~\ref{fig:global}. We adopt in this
section the notation of refs.~\cite{snoshow,snoten}.

\subsection{Predictions for the charged-current rate}
\label{subsec:ccrate}

The allowed range of neutrino parameters shown in
Figure~\ref{fig:global} corresponds to a range of predicted values for
$[{\rm CC}]_{\rm SNO}$, the to-be-measured SNO charged-current rate
divided by the predicted standard model rate for SNO charged-current
reactions. 

Figure~\ref{fig:cc} shows for each of the oscillation solutions the
predicted range allowed at a nominal $99.7$\% C.L. . Since the
predicted rate divided by the standard model rate depends upon the
survival probability of solar $\nu_e$'s as a function of energy, the
predicted values of $[{\rm CC}]_{\rm SNO}$ depend upon the recoil electron
energy threshold. Figure~\ref{fig:cc} gives results for both a $5$ MeV
threshold and an $8$ MeV threshold. The dashed error bar labeled
``Measure $3\sigma$'' represents the uncertainty in interpreting the
measurements according to the best available estimates~\cite{snoten},
which include the energy resolution, energy scale, $^8$B neutrino
energy spectrum, neutrino cross section, and counting statistics (for
one year of operation).

The numerical range for the [CC] ratio is, for a $5$ MeV 
 threshold: LMA ($0.20-0.41$), SMA ($0.34-0.49$), LOW ($0.36-0.42$),
 Just So$^2$ ($0.46-0.49$), VAC ($0.39-0.44$) for active neutrinos and SMA
 ($0.45-0.53$), Just So$^2$ ($0.45-0.50$) and VAC ($0.41-0.45$) for
 sterile neutrinos. For an $8$ MeV threshold, we find for [CC]: LMA
 ($0.20-0.41$), SMA ($0.40-0.51$), LOW ($0.36-0.42$), Just So$^2$
 ($0.46-0.49$), VAC ($0.29-0.41$) for active neutrinos and SMA
 ($0.45-0.58$), Just So$^2$ ($0.45-0.50$) and VAC ($0.33-0.39$) for
 sterile neutrinos.

For most of the currently allowed neutrino solution space, the
predicted value of $[{\rm CC}]_{\rm SNO}$ is expected to lie
reasonably close to the non-oscillation value of $[{\rm CC}]_{\rm SNO}
= 0.475$, which applies if neutrino oscillations do not occur and
Super-Kamiokande is measuring a pure solar $\nu_e$ beam.  The SMA and
Just So$^2$ active neutrino solutions, as well as the SMA and Just
So$^2$ sterile neutrino solutions, all predict charged-current rates
that are similar to the non-oscillation value.  Only for certain LMA
solution parameters is the predicted $[{\rm CC}]_{\rm SNO}$ rate well
separated from the Super-Kamiokande value.

The general trends shown in Figure~\ref{fig:cc} can be understood
quantitatively by a simple relation that is easily derived:

\begin{equation}
[{\rm CC}] = \frac{1}{1 - r} \times [R_{SK} - r f_B]
\times \frac{P_{\rm SNO}}{P_{\rm SK}} .
\label{eq:cc}
\end{equation}
Here, $R_{\rm SK}$ is the ratio ($0.475$) of the neutrino-electron
scattering rate observed by Super-Kamiokande to the rate expected on
the basis of the standard solar model, $r \sim 0.16$ is the ratio of
neutrino-electron scattering cross sections for muon and electron
neutrinos, and $f_{\rm B}$ is the ratio of the total $^8$B neutrino
flux to the standard solar model flux. The average survival
probabilities, $P_{\rm SNO}$ and $P_{\rm SK}$, refer to the energy
ranges most important for the SNO and the Super-Kamiokande
measurements.  Equation~\ref{eq:cc} is valid for solutions like the
LMA and LOW solutions (and somewhat less precisely for the SMA
solution) in which the survival probability is practically constant
over the region of interest.  For the LMA and LOW solutions
$P_{SNO}/P_{SK} \approx 1$ independent of energy thresholds.  The
derivation of Equation~\ref{eq:cc} neglects the small
energy-dependence of $r$.  

In addition to providing insight into the trends shown in
Figure~\ref{fig:cc}, Equation~\ref{eq:cc} can be used to make
`sanity-checks' of detailed numerical calculations. The reader can
make consistency checks of the results presented in
Figure~\ref{fig:cc} by using the data given in
Table~\ref{tab:bestfits} and Table~\ref{tab:fluxranges}.

\subsection{The ratio of neutral-current rate to charged-current rate}
\label{subsec:ncccratio}

\FIGURE[!h]{
\caption{The ratio of neutral-current rate to charged-current rate.
Figure~\ref{fig:double}a shows, for a 5 MeV threshold for the CC
measurement, the predicted double ratio of neutral-current rate to
charged-current rate for different neutrino scenarios.
Figure~\ref{fig:double}b shows the same ratio but for an 8 MeV CC
threshold.  The solid error bars shown represent the $99.73$\%
C.L. for the allowed regions of the eight currently favored neutrino
oscillation solutions in Figure~\ref{fig:global}.  The first five
solutions (from the left) refer to active neutrinos and the three
following solutions refer to sterile neutrinos.
\label{fig:double}}
\centerline{\psfig{figure=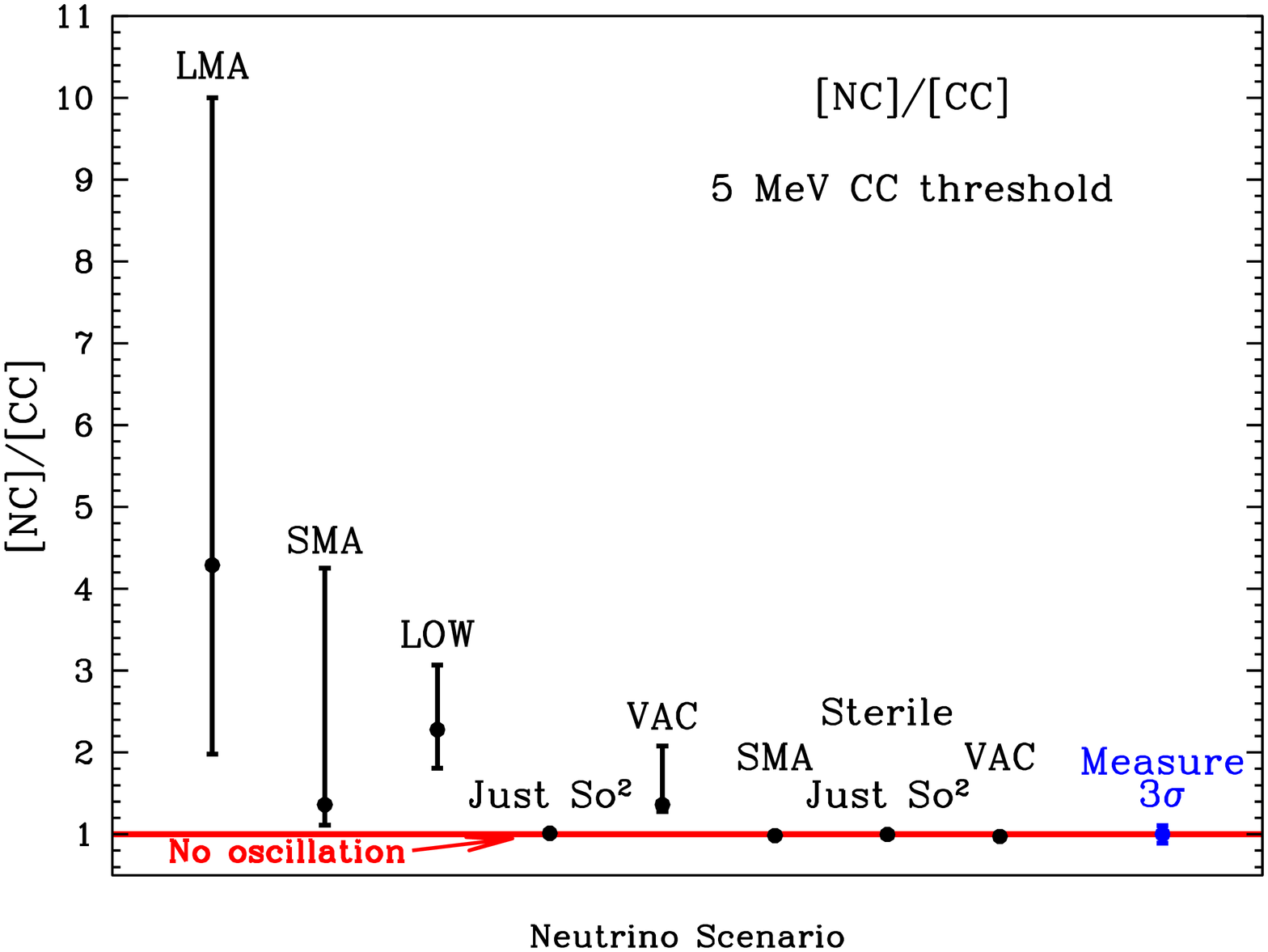,width=4.5in,angle=270}}
\centerline{\psfig{figure=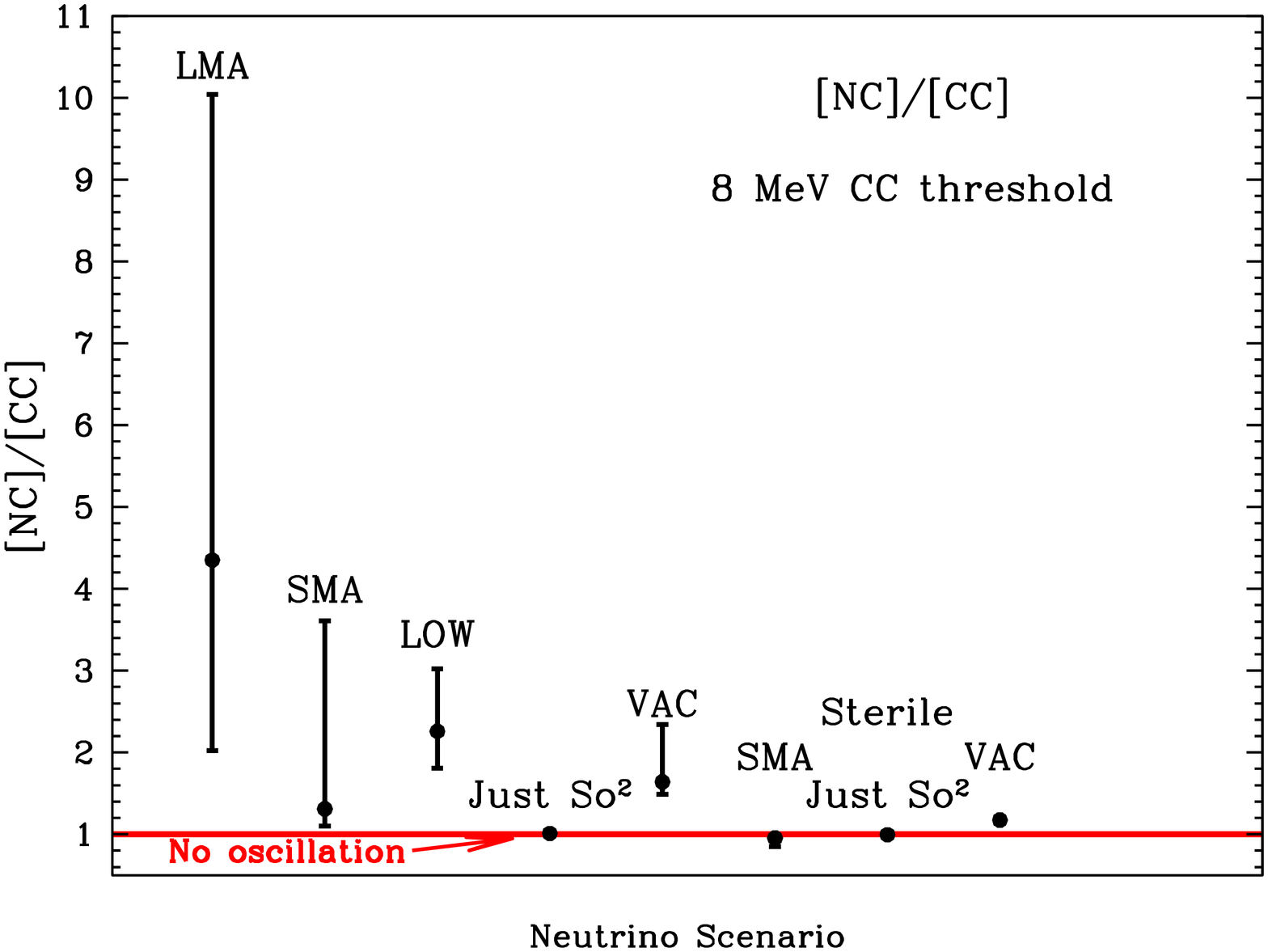,width=4.5in,angle=270}}
}

Figure~\ref{fig:double} shows the predicted values of the double
ratio, [NC]/[CC]. Here [NC]/[CC] is the ratio of the observed
neutral-current rate to the charged-current rate in SNO divided by the
same ratio calculated with the undistorted BP2000 fluxes.  The
standard model value for [NC]/[CC] is $1.0$.  Figure~\ref{fig:double}a
shows, for a 5 MeV threshold for the CC measurement, the predicted
double ratio of neutral-current to charged-current for the currently
allowed neutrino oscillation scenarios.  Figure~\ref{fig:double}b
shows the same ratio but for an 8 MeV CC threshold.  The solid error
bars shown represent the $99.73$\% C.L. for the allowed regions of the
six currently favored neutrino oscillation solutions in
Figure~\ref{fig:global}.  The error bar labeled ``Measure $3\sigma$''
represents the uncertainty in interpreting the measurements according
to the best available estimates~\cite{sno,snoten}, which include the
energy resolution, energy scale, $^8$B neutrino energy spectrum,
neutrino cross section, and counting statistics (for $5000$ CC
events).

The numerical range for the ratio [NC]/[CC] is, for a $5$ MeV CC
threshold: LMA ($2.0-10.0$), SMA ($1.1-4.0$), LOW ($1.8-3.1$),
Just So$^2$ ($1.011-1.016$) for active neutrinos and SMA ($0.964-0.997$)
and Just So$^2$ ($0.997-0.999$) for sterile neutrinos. For an $8$ MeV
CC threshold, we find for[NC]/[CC]: LMA ($2.05-10.05$), SMA
($1.1-3.4$), LOW ($1.8-3.0$), Just So$^2$ ($1.008-1.013$) for
active neutrinos and SMA ($0.89-0.99$) and Just So$^2$ ($0.993-0.997$)
for sterile neutrinos.

The numerical range for the ratio [NC]/[CC] is, for a $5$ MeV CC
 threshold: LMA ($2.0-10.0$), SMA ($1.1-4.2$), LOW ($1.8-3.1$),
 Just So$^2$ ($1.011-1.016$) and VAC ($1.3-2.1$)  for active neutrinos
 and SMA ($0.952-0.996$), Just So$^2$ ($0.997-0.999$) and VAC
 ($0.96-0.98$) for sterile neutrinos. For an $8$ MeV
 CC threshold, we find for[NC]/[CC]: LMA ($2.0-10.0$), SMA
 ($1.1-3.6$), LOW ($1.8-3.0$), Just So$^2$ ($1.008-1.012$) and VAC
 ($1.5-2.3$) for  active neutrinos and SMA ($0.86-0.99$), Just So$^2$
 ($0.993-0.997$) and VAC ($1.13-1.21$) for sterile neutrinos.

The LMA and LOW solutions are predicted to be well separated from the
non-oscillation value of [NC]/[CC] = 1.0. However, the Just So$^2$,
Sterile, and part of the SMA solution space are  practically coincident
with the no oscillation value.

The most striking way that Figure~\ref{fig:cc} and
Figure~\ref{fig:double} differ from our previous
results~\cite{snoten,snoshow} is that the Just So$^2$ are shown in the
newer results. The fact that $^8$B is treated as a free parameter in
the present analysis both allows the Just So$^2$ solutions to appear
and also decreases somewhat the predicted differences between the MSW
active neutrino solutions and the no-oscillation expectations.

The trends in the double ratio can be represented by an analytic
formula that is similar to, and derived in the same way, as
Equation~\ref{eq:cc} and which uses the same notation:
\begin{equation}
\frac{[NC]}{[CC]} = \frac{f_{\rm B}(1 - r)} 
{ [R_{SK} - r f_B]} \frac{P_{\rm SK}}{P_{\rm SNO}} .
\label{eq:nccc}
\end{equation}

\section{Discussion}
\label{sec:discussion}

For both active and for sterile neutrinos, we have obtained a global
solution, shown in Figure~\ref{fig:global}, for the eight allowed
regions of neutrino oscillation parameters.  

We allow the $^8$B and $hep$ neutrino fluxes created in the sun to be
free parameters, treating the fluxes consistently in both the fits to
the recoil energy spectrum and to the total event rates. However, we
updated input data from the BP2000 standard solar model including the
production profiles of the different neutrino sources, the number
density profiles for scatterers of active and of sterile neutrinos, as
well as the calculated fluxes, and their uncertainties, for all the
neutrino fluxes except the $^8$B and $hep$ fluxes. So, our analysis is
only a modest first step toward studying neutrino oscillations
independently of solar models. More experimental data are required
before one can begin to make studies of solar neutrinos that are truly
independent of solar models.

Six of the currently allowed regions are robustly allowed, i. e., the
LMA, SMA, LOW, and Just So$^2$ solutions for active neutrinos and the
SMA and Just So$^2$ solutions for sterile neutrinos, are essentially
unaffected by making common variations in the theoretical
analysis. The vacuum solutions at $\Delta m^2 \sim 10^{-10} {\rm ~
{\rm eV}^2}$ are rather fragile; whether or not they are present depends
upon how strongly one emphasizes the Super-Kamiokande spectral energy
data (see Section~\ref{sec:variations}). 

The Just So$^2$ solution with $\Delta m^2 \sim 6 \times 10^{-12}~{\rm
eV^2}$ is allowed in the present analysis because we treat the $^8$B
flux as a free parameter in fitting both the spectral and the total
rate data. The total $^8$B neutrino flux required for the Just So$^2$
solution is $3.3\sigma$ below the best-estimate $^8$B flux of the
standard solar model, using both the flux and the uncertainty of the
BP2000 model.

The Just So$^2$ solution, discussed in Section~\ref{sec:justso} and in
refs.~\cite{raghavan,KP96} and illustrated in
Figure~\ref{fig:spectrum}, describes in an obvious way all of the
solar neutrino results measured so far.  One can see immediately from
Figure~\ref{fig:spectrum} that the predicted distortion of the $^8$B
neutrino spectrum is very small in the region accessible to
Super-Kamiokande and SNO (above $5$ MeV). The day-night effect is
predicted to be zero. The rates measured in the radiochemical
experiments, chlorine and gallium, are accounted for by the strongly
suppressed $^7$Be $\nu_e$ flux, the only slightly suppressed $p-p$
$\nu_e$ flux ($\sim 23$\% for the gallium experiments), and the
inferred relatively low total $^8$B neutrino flux, $0.47$ of the
BP2000 value.

Unfortunately, the Just So$^2$ solution will not be distinguishable by
SNO from the no oscillation hypothesis (see Figure~\ref{fig:cc} and
Figure~\ref{fig:double}). BOREXINO and other experiments with
sensitivity below $1$ MeV will be required to identify Just So$^2$
oscillations if Nature has chosen this simple but elusive solution.

Figure~\ref{fig:cc} shows that the [CC] measurement by SNO will not
reveal strong evidence for neutrino oscillations unless Nature has
chosen a favorable part of the currently allowed LMA oscillation space
(cf. Figure~\ref{fig:global}).  The predictions for [CC] based upon
the best-fit parameters of four solutions, the active and sterile SMA
solutions and the active and sterile Just So$^2$ solutions, all lie
within the no-oscillation band illustrated in Figure~\ref{fig:cc}. The
fragile vacuum solutions with $\Delta m^2 \sim 10^{-10} {\rm ~ eV^2}$
both lie close to the no-oscillation band. Of the eight solutions
illustrated in Figure~\ref{fig:cc}, only the LMA solution offers the
possibility of a definitive ($> 3\sigma$) deviation from the no-oscillation
hypothesis.

The diagnostic power of the ratio of neutral-current rate to
charged-current rate, [NC]/[CC], is much greater. The current best
global solution predicts a significant deviation from the
no-oscillation hypothesis if either of the LMA, SMA, LOW or VAC
solutions for active neutrinos is valid. But the Just So$^2$ active
neutrino solution and the Just So$^2$ and SMA sterile neutrino
solutions predict a double ratio that can be consistent with the
no-oscillation value. The predicted numerical range for the [NC]/[CC]
ratio is given in Section~\ref{sec:sno} for each of the currently
allowed oscillation regions.

JNB and PIK acknowledge support from NSF grant No. PHY0070928 and PIK
acknowledges support from DOE grant DE-FG02-84ER40163. We are grateful
to A. Friedland and E. Lisi for valuable comments and suggestions.

\vfill\eject

\end{document}